\pdfoutput=1
\documentclass[acmsmall, nonacm]{acmart}

\usepackage{tikz}
\usetikzlibrary{arrows.meta, positioning, calc, decorations.pathreplacing, shapes}
\usepackage{listings}
\usepackage{mdframed}
\usepackage{booktabs}

\lstdefinelanguage{Dafny}{
  morekeywords={method, function, predicate, class, module, import, export,
    var, const, type, datatype, codatatype, newtype, trait, extends,
    returns, requires, ensures, modifies, reads, decreases, invariant,
    if, else, while, for, match, case, return, yield, break, continue,
    assert, assume, expect, print, reveal, calc,
    forall, exists, in, ghost, static, abstract, opaque,
    true, false, null, this, old, fresh, allocated,
    set, seq, multiset, map, iset, imap, array, string, int, nat, real, bool, char},
  sensitive=true,
  morecomment=[l]{//},
  morecomment=[s]{/*}{*/},
  morestring=[b]",
}

\mdfdefinestyle{keyinsights}{
  backgroundcolor=blue!5,
  linecolor=blue!40,
  linewidth=1pt,
  roundcorner=4pt,
  innertopmargin=10pt,
  innerbottommargin=10pt,
}

\mdfdefinestyle{sidebar}{
  backgroundcolor=gray!5,
  linecolor=gray!40,
  linewidth=1pt,
  roundcorner=4pt,
  innertopmargin=10pt,
  innerbottommargin=10pt,
}

\mdfdefinestyle{callout}{
  backgroundcolor=orange!5,
  linecolor=orange!50,
  linewidth=2pt,
  topline=false,
  bottomline=false,
  rightline=false,
  innertopmargin=6pt,
  innerbottommargin=6pt,
  innerleftmargin=12pt,
}

\title{Intent Formalization: A Grand Challenge for Reliable Coding in the Age of AI Agents}

\author{Shuvendu K. Lahiri}
\affiliation{
  \institution{Microsoft Research}
  \city{Redmond}
  \state{WA}
  \country{USA}
}
\email{shuvendu@microsoft.com}

\begin{abstract}
Agentic AI systems can now generate code with remarkable fluency, but a fundamental question remains: \emph{does the generated code actually do what the user intended?}
The gap between informal natural language requirements and precise program behavior---the \emph{intent gap}---has always plagued software engineering, but AI-generated code amplifies it to an unprecedented scale.
This article argues that \textbf{intent formalization}---the translation of informal user intent into a set of checkable formal specifications---is the key challenge that will determine whether AI makes software more reliable or merely more abundant.
Intent formalization offers a tradeoff spectrum suitable to the reliability needs of different contexts: from lightweight tests that disambiguate likely misinterpretations, through full functional specifications for formal verification, to domain-specific languages from which correct code is synthesized automatically.
The central bottleneck is \emph{validating specifications}: since there is no oracle for specification correctness other than the user, we need semi-automated metrics that can assess specification quality with or without code, through lightweight user interaction and proxy artifacts such as tests.
We survey early research that demonstrates the \textbf{potential} of this approach: interactive test-driven formalization that improves program correctness, AI-generated postconditions that catch real-world bugs missed by prior methods, and end-to-end verified pipelines that produce provably correct code from informal specifications.
We outline the open research challenges---scaling beyond benchmarks, achieving compositionality over changes, metrics for validating specifications, handling rich logics, designing human-AI specification interactions---that define a research agenda spanning AI, programming languages, formal methods, and human-computer interaction.
\end{abstract}

\begin{document}
\maketitle

% --- Key Insights Sidebar ---
\begin{mdframed}[style=keyinsights]
\textbf{Key Insights}
\begin{itemize}
  \item AI-generated code is plausible by construction but not correct by construction. The \emph{intent gap}---the distance between what a user means and what a program does---is the central reliability bottleneck.
  \item \textbf{Intent formalization}---automatically translating informal user intent into checkable specifications---can close this gap. It offers a tradeoff spectrum suited to different reliability needs: from lightweight tests, through full functional specifications for formal verification, to domain-specific languages from which correct code is synthesized automatically.
  \item The central bottleneck is \textbf{validating specifications}: since there is no oracle for specification correctness other than the user, we need semi-automated metrics that assess specification quality with or without code, through lightweight user interaction and proxy artifacts such as tests.
  \item Early research demonstrates the \textbf{potential}---improved developer correctness, new bugs caught, and verified pipelines from informal prose to correct code---but key \textbf{challenges} remain: scaling beyond benchmark problems, validating specifications without an oracle, and integrating intent formalization into agentic developer workflows.
  \item Sustained research investment is needed across AI, programming languages, formal methods, and HCI to scale intent formalization to production systems.
\end{itemize}
\end{mdframed}

% =============================================================================
\section{Introduction}
\label{sec:intro}

\emph{Vibe coding}---a term coined by Andrej Karpathy~\cite{karpathy2025vibe}---captures the new reality of AI-powered software development: developers describe what they want in natural language, accept AI-generated code with minimal or no review, and ``forget that the code even exists.''
A growing ecosystem of agentic coding tools---GitHub Copilot coding agent, Claude Code, and others---now synthesizes entire functions, modules, and systems from brief prompts, planning, writing code, running tests, and iterating---often autonomously.
This represents the purest manifestation of the intent gap: the user has \emph{intent} but never inspects the \emph{implementation}, relying entirely on the AI to bridge the two.

But a fundamental question remains unanswered: \emph{does the generated code actually do what the user intended?}

Consider a simple request: ``given a list of integers, remove duplicates.''
Does this mean keep one copy of each element (e.g., \texttt{[1,2,3,2,4]} $\to$ \texttt{[1,2,3,4]})?
Or does it mean remove all elements that appear more than once, keeping only unique ones (e.g., \texttt{[1,2,3,2,4]} $\to$ \texttt{[1,3,4]})?
A human developer resolves such ambiguities through domain knowledge and conversation.
An LLM resolves them through statistical pattern-matching against training data---with no grounding in the user's specific intent.
The result is code that \emph{looks right} but may silently deviate from what the user actually wanted.

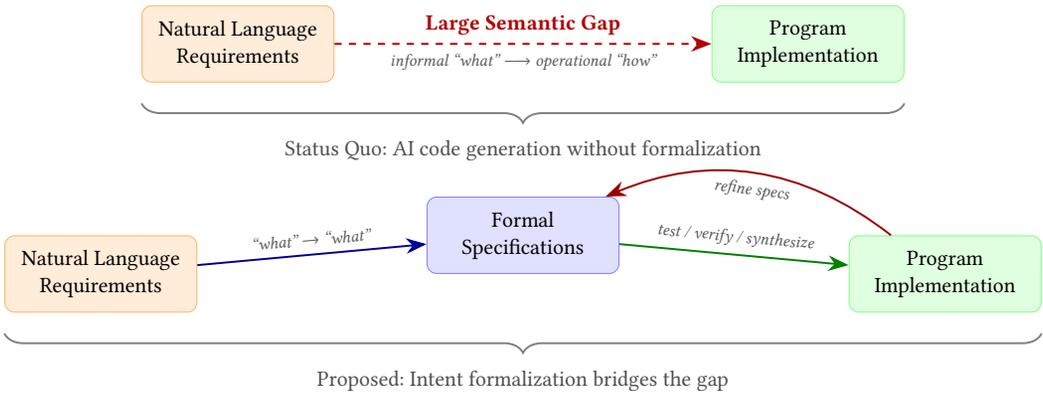
\begin{figure}[t]
\centering
\resizebox{\columnwidth}{!}{%
\begin{tikzpicture}[
  node distance=1.8cm and 2.5cm,
  box/.style={draw, rounded corners=4pt, minimum width=2.8cm, minimum height=1.1cm, align=center, font=\small},
  nlbox/.style={box, fill=orange!15, draw=orange!60},
  specbox/.style={box, fill=blue!12, draw=blue!50},
  codebox/.style={box, fill=green!12, draw=green!50},
  arrow/.style={-{Stealth[length=3mm]}, thick},
  dasharrow/.style={-{Stealth[length=3mm]}, thick, dashed, red!70!black},
  label/.style={font=\scriptsize\itshape, text=gray!70!black},
]

% Top row: the traditional (large) gap
\node[nlbox] (nl) {Natural Language\\Requirements};
\node[codebox, right=5.5cm of nl] (code) {Program\\Implementation};

% Big red dashed arrow showing the gap
\draw[dasharrow] (nl) -- node[above, font=\small\bfseries, text=red!70!black] {Large Semantic Gap} node[below, label] {informal ``what'' $\longrightarrow$ operational ``how''} (code);

% Bottom row: with specifications
\node[specbox, below=2.2cm of $(nl)!0.5!(code)$] (spec) {Formal\\Specifications};

\node[nlbox, below left=2.2cm and 2.0cm of nl.south east] (nl2) {Natural Language\\Requirements};
\node[codebox, below right=2.2cm and 2.0cm of code.south west] (code2) {Program\\Implementation};

\draw[arrow, blue!60!black] (nl2) -- node[above, label, sloped] {``what'' $\to$ ``what''} (spec);
\draw[arrow, green!50!black] (spec) -- node[above, label, sloped] {test / verify / synthesize} (code2);
\draw[arrow, red!60!black, bend right=30] (code2) to node[below, label, sloped] {refine specs} (spec);

% Brace and labels
\draw[decorate, decoration={brace, amplitude=8pt, mirror}, thick, gray] ([yshift=-0.3cm]nl.south west) -- ([yshift=-0.3cm]code.south east) node[midway, below=0.4cm, font=\small, text=gray!60!black] {Status Quo: AI code generation without formalization};

\draw[decorate, decoration={brace, amplitude=8pt, mirror}, thick, gray] ([yshift=-0.3cm]nl2.south west) -- ([yshift=-0.3cm]code2.south east) node[midway, below=0.4cm, font=\small, text=gray!60!black] {Proposed: Intent formalization bridges the gap};

\end{tikzpicture}
}%
\caption{The intent gap in software development. \textbf{Top:} Traditional AI code generation translates informal, ambiguous natural language directly into operational code, leaving a large semantic gap. \textbf{Bottom:} Formal specifications serve as an intermediate ``what'' layer, reducing the gap and enabling enforcement through testing and verification.}
\label{fig:intent-gap}
\end{figure}

For example, a typical LLM response to this prompt produces:

\begin{lstlisting}[language=Python, caption={Plausible but wrong if the user intended to remove all numbers that have duplicates.}]
def remove_duplicates(numbers):
    return list(dict.fromkeys(numbers))
\end{lstlisting}

This keeps one copy of each element in order: \texttt{remove\_duplicates([1,2,3,2,4])} returns \texttt{[1,2,3,4]}.
But a user who meant ``remove all numbers that appear more than once''~\cite{endres2024fse} expected \texttt{[1,3,4]}---the \texttt{2} should be gone entirely.
A formal postcondition disambiguates the intent:

\begin{lstlisting}[language=Python, caption={A postcondition capturing the ``remove elements with duplicates'' intent.}, numbers=none]
assert all(numbers.count(x) == 1 for x in result)
assert all(x in result for x in numbers if numbers.count(x) == 1)
\end{lstlisting}

The ambiguity in this simple example illustrates a broader phenomenon.
The \emph{intent gap}---the semantic distance between what a user means and what a program does---has always existed in software engineering (Figure~\ref{fig:intent-gap}), but AI amplifies it in two ways:

% =============================================================================
% Figure 1: The Intent Gap — from NL requirements to code
% =============================================================================

\begin{enumerate}
  \item \textbf{Scale without scrutiny.} AI generates code faster than humans can review it. The ratio of code produced to code carefully examined is growing rapidly, and traditional safeguards---code review, manual testing---cannot keep pace.
  \item \textbf{Plausibility without correctness.} LLM-generated code is \emph{plausible by construction}---it looks right, compiles, and often passes a few tests---but it is not \emph{correct by construction}. Subtle errors hide behind surface-level fluency, making AI-generated bugs harder to spot than hand-written ones.
\end{enumerate}

We argue that the path to reliable AI-generated code is not better code generation---it is \textbf{intent formalization}: the automatic translation of informal user intent into checkable formal specifications.
Rather than asking ``can AI write the code?'' we should ask ``can AI help us specify what the code should do---and then verify that it does?''

\begin{mdframed}[style=callout]
Intent formalization offers a tradeoff spectrum suited to different reliability needs.
At one end, lightweight specifications including tests target \emph{points of likely ambiguity}---the places where different LLMs or agents would generate semantically different code from the same prompt---acting as cost-effective \textbf{guardrails}.
In the middle, full functional specifications in verification-aware languages like Dafny, F*, and Verus enable machine-checked proofs of correctness.
At the far end, domain-specific languages (DSLs) serve as complete specifications from which provably correct code is generated automatically.
\end{mdframed}

This article presents a framework for intent formalization, surveys early research that demonstrates its promise, and outlines the open problems that define a research agenda for the next decade.
The goal is to make a case for sustained investment by the research community and industry in intent formalization as a \textbf{first-class research priority}.

\subsection{Why Now?}
\label{sec:why-now}

The need for intent formalization was recognized with the advent of the first generation of neural code-generation models (TiCoder~\cite{lahiri2022ticoder}).
OpenAI Codex (back in 2021) demonstrated that LLMs could produce syntactically correct code from natural language prompts---but also revealed how easily plausible code could silently deviate from the user's intent.
At that stage, however, AI coding tools were primarily \emph{autocomplete assistants}: they suggested individual lines or small blocks of code, and the developer retained full control, reviewing each suggestion before accepting it.
The human remained the primary author; the AI was a productivity aid.

Since then, the landscape has transformed dramatically.
Agentic coding tools autonomously write, test, and debug entire features.
Vibe coding~\cite{karpathy2025vibe} encourages developers to describe intent and ``let the AI handle it.''
The human has shifted from \emph{author} to \emph{supervisor}---and in many vibe coding scenarios, to a passive \emph{consumer} of AI-generated code.

This shift has three consequences that make intent formalization critical:
\begin{enumerate}
  \item \textbf{Human review is being bypassed.} When developers accept AI-generated code with minimal or no review, the traditional human safeguard against intent violations disappears.
  Formal specifications become the \emph{only} scalable mechanism for checking that generated code matches user intent.
  \item \textbf{The attack surface has exploded.} AI-generated code is entering production at unprecedented scale---including safety-critical and security-sensitive systems.
  A single specification gap in a parser, an authentication module, or a financial transaction handler can have outsized consequences.
  \item \textbf{The technology is ready.} The same LLMs that power vibe coding can also generate specifications.
  Verification infrastructure (SMT solvers, proof assistants, type systems) has matured over decades.
  For the first time, it is feasible to close the loop: generate code, generate specifications, and verify one against the other---all within an AI-assisted workflow.
\end{enumerate}

% =============================================================================
\section{What Is Intent Formalization?}
\label{sec:what}

\begin{figure}[t]
\centering
\resizebox{\columnwidth}{!}{%
\begin{tikzpicture}[
  node distance=0.8cm and 0.6cm,
  layer/.style={draw, rounded corners=5pt, minimum height=1.2cm, align=center, font=\small},
  arrow/.style={-{Stealth[length=2.5mm]}, thick, gray},
  verif/.style={draw, dashed, rounded corners=3pt, fill=green!5, draw=green!50, font=\scriptsize, align=center, minimum height=0.8cm},
]

% Spectrum bar (specifications)
\fill[left color=orange!20, right color=blue!20] (0,0) rectangle (18,0.5);
\node[font=\small\bfseries, anchor=west] at (0.2,0.25) {Partial};
\node[font=\small\bfseries, anchor=east] at (17.8,0.25) {Complete};

% Specification layers
\node[layer, fill=orange!10, draw=orange!40, minimum width=3.5cm] (tests) at (2.3, 1.8) {\textbf{Tests}\\(I/O examples)};
\node[layer, fill=yellow!10, draw=yellow!50, minimum width=3.8cm] (contracts) at (6.8, 1.8) {\textbf{Code Contracts}\\(assertions/pre/post/invariants)};
\node[layer, fill=blue!10, draw=blue!50, minimum width=3.5cm] (logical) at (11.3, 1.8) {\textbf{Logical Contracts}\\(Dafny/Verus)};
\node[layer, fill=purple!10, draw=purple!50, minimum width=3.5cm] (dsls) at (15.8, 1.8) {\textbf{DSLs}\\(full specifications)};

% Tools below each
\node[font=\scriptsize, text=gray!70!black, align=center] at (2.3, 0.75) {TiCoder};
\node[font=\scriptsize, text=gray!70!black, align=center] at (6.8, 0.75) {nl2postcond\\ClassInvGen};
\node[font=\scriptsize, text=gray!70!black, align=center] at (11.3, 0.75) {Auto-Verus\\VeriStruct};
\node[font=\scriptsize, text=gray!70!black, align=center] at (15.8, 0.75) {3DGen};

% Connecting arrows
\draw[arrow] (tests) -- (contracts);
\draw[arrow] (contracts) -- (logical);
\draw[arrow] (logical) -- (dsls);

% Top label
\node[font=\small\itshape, text=gray] at (9, 3.0) {Increasing correctness coverage $\longrightarrow$};

% Verification row below spectrum bar
\node[verif, minimum width=7.5cm] (dynamic) at (4.55, -0.7) {\textbf{Dynamic checking}: tests, runtime contracts};
\node[verif, minimum width=3.5cm] (static) at (11.3, -0.7) {\textbf{Static verification}: proofs};
\node[verif, minimum width=3.5cm] (synth) at (15.8, -0.7) {\textbf{Verified synthesis}: code generation};

\end{tikzpicture}
}%
\caption{The spectrum of intent formalization. \textbf{Specifications} (top row) range from partial (tests) to complete (domain-specific languages)---all are formal and checkable, differing in correctness coverage. Tests and code contracts are checked \emph{dynamically}; logical contracts require \emph{static verification} via a program verifier; DSLs enable \emph{verified synthesis} of correct-by-construction code. LLMs can help generate artifacts at every level.}
\label{fig:spectrum}
\end{figure}
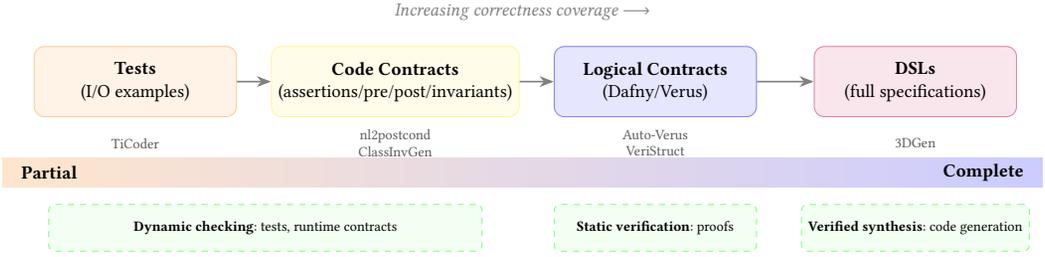

We define \textbf{intent formalization} as the problem of automatically translating informal user intent into a set of formal, checkable program specifications.
The resulting specifications span a spectrum of increasing expressiveness (Figure~\ref{fig:spectrum}):

% =============================================================================
% Figure: The Spectrum of Intent Formalization
% =============================================================================

\begingroup\emergencystretch=2em
\begin{itemize}
  \item \textbf{Tests} (input/output examples): concrete behavioral expectations, e.g.,
  {\small\texttt{remove\_duplicates\hspace{0pt}([1,\hspace{0pt}2,\hspace{0pt}3,\hspace{0pt}2,\hspace{0pt}4])}} should return {\small\texttt{[1,3,4]}}.
  \item \textbf{Code contracts} (assertions, pre/\allowbreak postconditions, invariants): executable specifications checked dynamically at runtime, from inline assertions such as
  {\small\texttt{assert all(\hspace{0pt}nums.count(x)\hspace{0pt}==1 for x in res)}} to function-level postconditions and class invariants.
  \item \textbf{Logical contracts}: specifications in verification-aware languages such as Dafny~\cite{leino2010dafny}, F$^*$~\cite{swamy2016fstar}, and Verus~\cite{lattuada2024verus} that use quantifiers, ghost variables, and recursive predicates.
  These require a \emph{program verifier} (e.g., an SMT solver) to check statically for all possible inputs.
  \item \textbf{Domain-specific languages (DSLs)}: complete formal specifications in a specialized notation from which provably correct code is generated automatically via verified compilation or synthesis.
\end{itemize}
\endgroup

Crucially, intent formalization is \emph{not} restricted to verification-aware languages---tests and code contracts apply to any mainstream language (Python, Java, C++, Rust).
Verification-aware languages such as Dafny and Verus can offer stronger guarantees through machine-checked proofs, and domain-specific languages go further still---serving as complete specifications from which correct code is synthesized automatically.
The first two levels of the spectrum, however, are language-agnostic and immediately deployable.

Tests and code contracts can be checked \emph{dynamically}---by running the program.
Logical contracts require \emph{static verification} via a program verifier, offering stronger guarantees but demanding more sophisticated specifications and tooling.
DSLs occupy the far end of the spectrum: the specification is complete enough that correct code is \emph{synthesized} automatically via verified compilation (Section~\ref{sec:3dgen}).
A key insight is that these levels are not alternatives---they are \emph{complementary}, and progress at any level enables progress at the others.
For instance, tests can validate postconditions, postconditions can guide invariant discovery, and invariants can anchor full proofs.\footnote{Intent formalization is not just limited to code generation. Even a \emph{test oracle}---the expected output for a single test input---is an instance of intent formalization~\cite{dinella2022toga}.}

\begin{mdframed}[style=callout]
\textbf{Two important distinctions.}
\emph{Not autoformalization.} Intent formalization is distinct from autoformalization---the translation of complete natural language specifications into formal logic~\cite{wu2022autoformalization}.
Autoformalization seeks full fidelity to a complete source text; intent formalization offers a cost-effective spectrum---from disambiguating the most ambiguity-prone properties of an inherently incomplete NL prompt to domain-specific languages from which correct code is synthesized automatically.\\[4pt]
\emph{Complementary to spec-driven development.} Tools like GitHub Spec Kit~\cite{github2025speckit} structure AI coding around natural language requirements, improving traceability.
But these specifications remain informal and uncheckable.
Intent formalization closes this gap by producing formal, checkable specifications that can be mechanically verified against generated code.
\end{mdframed}

\subsection{Specifications vs.\ Verification}
\label{sec:spec-vs-verif}

It is important to distinguish \emph{specifications} from \emph{verification}.
Specifications describe \emph{what} the code should do; verification checks that the code actually does it.
\begin{itemize}
  \item \textbf{Testing} verifies specifications against \emph{finitely many} inputs. It is lightweight and widely applicable, but inherently incomplete: passing all tests does not guarantee correctness on unseen inputs.
  \item \textbf{Runtime checks} verify specifications on \emph{every execution} in production. Postconditions and assertions are evaluated at runtime, catching violations as they occur. This provides stronger coverage than offline testing but incurs runtime overhead and only detects errors when they are triggered.
  \item \textbf{Proofs} verify specifications against \emph{all possible} inputs. They are generated by SMT solvers~\cite{demoura2008z3} or proof assistants and provide mathematical guarantees, but require richer specifications and more sophisticated automation.
\end{itemize}

Intent formalization focuses on the \emph{specification side}: once specifications are in hand, existing verification infrastructure---test runners, SMT solvers, proof assistants---can check them.
\begin{mdframed}[style=callout]
A key bottleneck today is the absence of formal specifications to verify against.
While verification technology itself continues to advance, specifications are a prerequisite---verification tools remain idle without them.
Intent formalization provides a tradeoff spectrum: even targeted specifications at points of likely ambiguity serve as cost-effective guardrails~\cite{lahiri2022ticoder}, and teams can invest further---through full functional specifications for formal verification---as reliability needs demand.
\end{mdframed}

% =============================================================================
\section{Early Research on Intent Formalization}
\label{sec:evidence}

Intent formalization is not a speculative direction.
Multiple independent lines of early research---primarily on benchmark problems---provide concrete evidence of the promise of this approach.
We organize this evidence along four dimensions, progressing from individual capabilities to end-to-end systems.

\subsection{LLMs Can Generate Meaningful Specifications}
\label{sec:postcond}

LLMs prompted with natural language descriptions can generate \emph{postconditions}---executable assertions that constrain function outputs for arbitrary inputs~\cite{endres2024fse}.
On the Defects4J benchmark~\cite{just2014defects4j} (hundreds of real bugs across large Java projects), LLM-generated postconditions caught one in eight real bugs, including bugs missed by the classic Daikon invariant detector~\cite{ernst1999daikon}.
GPT-4 generated postconditions with substantially higher soundness and completeness scores than GPT-3.5 or CodeLlama, suggesting that specification quality scales with model capability.

Beyond function-level postconditions, ClassInvGen~\cite{sun2025classinvgen} synthesizes \emph{class invariants}---the key properties that hold an entire module together---for C++ data structures.
A single well-chosen class invariant can replace hundreds of function-level specifications, making review tractable and surfacing the crucial non-trivial properties that practitioners care about most.
ClassInvGen outperforms both direct LLM prompting and Daikon on this task.
VeriStruct~\cite{sun2026veristruct} scales further to entire data-structure modules in Verus, verifying nearly all functions across 11 modules including linked lists, hash maps, and B-trees.

These results establish that \emph{LLMs can produce specifications encoding real semantic understanding, not just syntactic patterns}.

But a prior question arises: how do we \emph{measure} whether generated specifications are good enough?

\subsection{Measuring Specification Quality}
\label{sec:uif}
\label{sec:metrics}

A fundamental challenge for intent formalization is that \emph{there is no oracle for specification correctness other than the user}.
Code can be tested against specifications, but who tests the specifications?
Unlike code, where test suites provide an independent check, a generated specification has no ground truth to compare against---the user's intent exists only in the user's head.
This makes \textbf{validating specifications} one of the foremost open problems.

We advocate for automated metrics grounded in two properties~\cite{endres2024fse, lahiri2024fmcad}:

\begin{itemize}
  \item \textbf{Soundness}: the specification is consistent with correct behavior---it does not reject valid implementations.
  \item \textbf{Completeness}: the specification is discriminating---it rejects incorrect implementations.
\end{itemize}

One approach~\cite{lahiri2024fmcad} operationalizes these properties using only \emph{tests} (input/output pairs), without requiring the code itself:
\begin{itemize}
  \item A specification $S$ is \emph{sound} with respect to a test suite $T$ if $S$ is satisfied on every test $(i, o) \in T$.
  \item A specification $S$ is \emph{complete} with respect to $T$ if, for each test $(i, o) \in T$, $S$ \emph{fails} when the output $o$ is replaced by a mutated output $o'$.
  To accommodate non-determinism (where multiple outputs may be valid for a given input), completeness is measured as the fraction of output mutations that the specification detects.
\end{itemize}

The main challenge for soundness is assembling an exhaustive set of tests: a specification that passes a handful of tests may still reject valid behavior on untested inputs.
Completeness is harder still: it requires not only that the specification capture \emph{exactly} what the user means, but also that the output mutations reflect \emph{natural} mistakes---the kinds of errors that humans, and increasingly AI, actually make.

Since both properties are defined in terms of evaluating a specification against tests, they can be checked before any implementation is written~\cite{endres2024fse, lahiri2024fmcad}.
For simple executable assertions this evaluation is straightforward, but specifications in verification-aware languages often involve quantifiers, recursive predicates, and ghost variables that cannot be directly executed.
Lahiri et al.~\cite{lahiri2024fmcad} address this by proposing novel symbolic techniques to evaluate such rich specifications against concrete test inputs, enabling automated soundness and completeness checking even for complex logical contracts.
This provides objective, reproducible measures of specification quality---just as NL-to-Code research advanced through test-based benchmarks like HumanEval~\cite{liu2023evalplus}.
Test-based evaluation is one promising approach; other techniques---symbolic analysis, property-based testing, and targeted user interaction---may further strengthen specification validation.

As an example of the power of automated metrics, consider a GPT-4--generated Dafny specification for ``common elements'' labeled ``strong'' by expert reviewers~\cite{misu2024fse}:

\begin{lstlisting}[language=Dafny, caption={A Dafny specification labeled ``strong'' by expert reviewers---but automated completeness metrics reveal it is incomplete~\cite{lahiri2024fmcad}.}, numbers=none, xleftmargin=0em]
ensures forall x :: x in result ==> (InArray(a, x) && InArray(b, x))
ensures forall i,j :: 0<=i<j<|result| ==> result[i] != result[j]
\end{lstlisting}

This says every element in \texttt{result} appears in both inputs, and results have no duplicates---\emph{sound}, but \emph{incomplete}.
Automated symbolic testing~\cite{lahiri2024fmcad} found that the implication (\texttt{==>}) should be a bi-implication (\texttt{<==>}): without it, the empty list trivially satisfies the specification.
The corrected version:

\begin{lstlisting}[language=Dafny, caption={Corrected specification with bi-implication.}, numbers=none]
ensures forall x :: x in result <==> (InArray(a, x) && InArray(b, x))
\end{lstlisting}

Across the full evaluation, automated metrics found three mislabeled and two inconsistent specifications introduced by copy-paste errors---all missed by human labeling~\cite{lahiri2024fmcad}.
\emph{You cannot improve what you cannot measure}, and automated metrics are essential for scaling intent formalization.

These metrics also enable downstream \emph{proof automation}.
The Auto-Verus system~\cite{chen2025iclr} uses soundness and completeness metrics~\cite{lahiri2024fmcad} to filter LLM-generated specifications and proofs for Rust/Verus programs, bootstrapping high-quality training data through a self-evolution cycle and achieving $3.6\times$ higher proof accuracy than GPT-4o zero-shot.
\begin{mdframed}[style=callout]
More abundant and higher-quality specifications can enhance AI's ability to generate proofs.
\end{mdframed}

With reliable metrics in hand, we can ask whether specifications improve code generation itself---not just evaluate it after the fact.
When the user \emph{must} be consulted, the interaction should be optimized to maximize the number of correct specifications obtained per user query---a principle embodied by TiCoder.

\subsection{Interactive Intent Formalization}
\label{sec:ticoder}

\begin{figure}[t]
\centering
\resizebox{\columnwidth}{!}{%
\begin{tikzpicture}[
  node distance=0.9cm and 1.4cm,
  box/.style={draw, rounded corners=3pt, minimum width=2.2cm, minimum height=0.85cm, align=center, font=\small},
  userbox/.style={box, fill=orange!15, draw=orange!60},
  aibox/.style={box, fill=blue!12, draw=blue!50},
  outputbox/.style={box, fill=green!12, draw=green!50},
  arrow/.style={-{Stealth[length=2.5mm]}, thick},
  backarrow/.style={-{Stealth[length=2.5mm]}, thick, dashed, gray},
]

% User icon (stick figure)
\node (user) {%
  \begin{tikzpicture}[scale=0.5]
    \draw[thick, orange!70!black] (0,1.8) circle (0.3);
    \draw[thick, orange!70!black] (0,1.5) -- (0,0.7);
    \draw[thick, orange!70!black] (-0.4,1.3) -- (0,1.1) -- (0.4,1.3);
    \draw[thick, orange!70!black] (-0.3,0) -- (0,0.7) -- (0.3,0);
  \end{tikzpicture}
};
\node[below=-0.1cm of user, font=\scriptsize\bfseries, text=orange!70!black] (ulabel) {Developer};

% Horizontal flow
\node[userbox, right=1.2cm of user] (nl) {NL\\Prompt};
\node[aibox, right=1.4cm of nl] (gen) {LLM generates\\candidate code};
\node[aibox, right=1.4cm of gen] (testgen) {Generate\\candidate tests};
\node[userbox, right=1.4cm of testgen] (approve) {Approve /\\reject tests};
\node[outputbox, right=1.4cm of approve] (result) {Ranked code +\\approved tests};

% Forward flow
\draw[arrow] (user) -- (nl);
\draw[arrow] (nl) -- (gen);
\draw[arrow] (gen) -- (testgen);
\draw[arrow] (testgen) -- (approve);
\draw[arrow] (approve) -- node[above, font=\scriptsize\itshape, text=gray] {prune} (result);

% Feedback loop — routed below
\draw[backarrow] (approve.south) -- ++(0,-0.8) -| node[below, pos=0.25, font=\scriptsize\itshape, text=gray] {iterate} (testgen.south);

% User participates in approval
\draw[arrow, orange!60, dashed] (user.east) -- ++(0.3,0) |- (approve.south west);

% Side annotation
\node[above=0.15cm of approve, font=\scriptsize, text=orange!70!black, align=center] {Yes / No / Undef};

\end{tikzpicture}
}%
\caption{The TiCoder interactive workflow for test-driven user-intent formalization. The developer provides a natural language prompt; the LLM generates candidate code and tests. The user iteratively approves or rejects tests, which prune and rank candidates. This improves correct evaluation of AI code from 40\% to 84\%~\cite{fakhoury2024tse}.}
\label{fig:ticoder}
\end{figure}
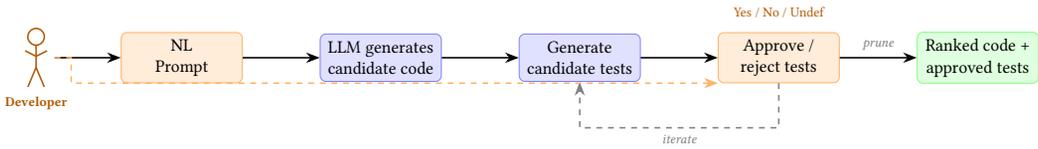

The TiCoder system~\cite{lahiri2022ticoder, fakhoury2024tse} uses intent formalization \emph{interactively} during code generation (Figure~\ref{fig:ticoder}).

% =============================================================================
% Figure 4: TiCoder Interactive Workflow (horizontal with user icon)
% =============================================================================

Rather than accepting whatever code an LLM produces, TiCoder generates candidate tests that \emph{prioritize points of ambiguity}---inputs where different code candidates produce different outputs---and asks the user to classify each as ``Yes,'' ``No,'' or ``Undef.''
Approved tests prune incorrect candidates via code execution.
This embodies the pay-as-you-go philosophy: TiCoder starts at the cost-effective end of the spectrum, targeting the tests most likely to expose where the LLM's interpretation diverges from the user's intent.

\begin{mdframed}[style=sidebar]
\textbf{Sidebar: A TiCoder Interaction}\\[4pt]
\textbf{Prompt:} ``Find the shared elements from two lists.''\\[4pt]
\textbf{Generated tests:}
\begin{enumerate}
  \item \texttt{common([1,2,3],[2,3,4]) == [2,3]} \hfill \textsf{User: Yes}
  \item \texttt{common([1,2,2],[2,2,3]) == [2,2]} \hfill \textsf{User: No}
  \end{enumerate}
By rejecting test~2, the user disambiguates that the result should be a \emph{set}, not a multiset.
TiCoder prunes candidates that preserve duplicates.
\end{mdframed}

A small-scale controlled study with 15 professional developers on code-generation benchmarks~\cite{fakhoury2024tse} found:
developers correctly evaluated AI-generated code roughly twice as often with TiCoder as without it ($p < 0.001$), cognitive load dropped significantly ($p = 0.007$), approved tests persisted as regression tests, and the majority of participants preferred TiCoder.
On these benchmarks, a modest amount of intent formalization---approving a few tests---yields a significant return in correctness.

\subsection{End-to-End Verified Pipelines}
\label{sec:3dgen}

At the far end of the spectrum, the 3DGen system~\cite{fakhoury2025icse} demonstrates intent formalization at the DSL level, where the specification is complete enough to generate code automatically.
3DGen uses a multi-agent AI architecture to translate informal RFC prose into formal specifications in the 3D domain-specific language, with symbolic test generation providing feedback for iterative refinement.
The verified 3D specifications compile via EverParse~\cite{swamy2022everparse} into provably correct, memory-safe C or Rust binary parsers---the specification \emph{is} the program, mediated by verified synthesis.
It has produced verified parsers for 20 standard network protocol formats (DNS, TLS extensions, QUIC)---demonstrating the potential of the full spectrum, from informal prose through a DSL to provably correct, deployable code.

% =============================================================================
\section{A Research Agenda}
\label{sec:challenges}

The early research above demonstrates promise on benchmark problems, but also reveals how far we are from a general solution.
Seven open problems define the research frontier, spanning AI capabilities, formal methods infrastructure, human interaction design, and software engineering practice.

\paragraph{From benchmarks to real-world systems.}
Current results target self-contained algorithmic functions.
Real-world software has side effects, mutable state, concurrency, and complex dependencies.
What does a ``postcondition'' mean for an asynchronous event handler or a machine learning pipeline?
We need benchmarks, metrics, and specification idioms for real-world intent formalization.

\paragraph{Change intent and compositionality.}
In practice, most software development involves \emph{changing} existing code, not writing from scratch.
Here, intent comprises not only the natural language change description---``fix this bug,'' ``add caching,'' ``handle the empty-input edge case''---but also the existing source code, tests, and specifications that define the current behavior.
Intent formalization must therefore capture what should change about behavior and compose with existing specifications.
A closely related challenge is \emph{code translation}---for example, migrating legacy C codebases to Rust.
Early work such as SpecTra~\cite{nitin2024spectra} shows that generating intermediate informal specifications from source code and using them to guide translation significantly improves correctness.
Formalizing the intent of code to be translated is crucial for reliable code translation.

\paragraph{Identifying what to clarify cost-effectively.}
For a practitioner, the value of a specification is measured by how many bugs it can prevent.
We therefore need metrics for \emph{ranking} specifications by their expected impact, especially when surfacing them to users for validation.
TiCoder takes a first step by generating diverse code candidates and targeting tests at inputs where candidates disagree---but this requires sampling the space of plausible implementations, which becomes expensive for large code blocks or multi-function tasks.
Scalable methods are needed to prioritize specifications by their bug-prevention value without exhaustively enumerating candidate programs.

\paragraph{Automated metrics for spec validation.}
Since there is no oracle for specification correctness other than the user, validating generated specifications is a foremost challenge.
Progress requires incorporating multiple complementary signals: tests and mutation analysis as automated proxies~\cite{endres2024fse, lahiri2024fmcad}, targeted user feedback to resolve ambiguities that automation cannot settle, and cross-checking across different artifacts such as code, docstrings, and formal annotations~\cite{sun2024clover}.

\paragraph{Rich logics and quantifiers.}
Verification-aware languages use quantifiers, recursive predicates, and ghost variables.
LLMs struggle with these constructs, and existing verifiers have fundamental limitations in unrolling recursive predicates for complex concrete test inputs, making automated soundness and completeness evaluation difficult~\cite{lahiri2024fmcad}.

\paragraph{Human-AI interaction for specification.}
TiCoder's approve/reject loop significantly improves correctness on benchmark problems, but real-world specification needs richer interaction: natural language explanations of formal properties, confidence-calibrated suggestions, and specification templates---a largely unexplored HCI design space.

\paragraph{Integration into developer workflows.}
Intent formalization must integrate naturally into modern developer workflows---from issue creation, where specifications capture intended behavior before code is written; through code review, where specifications surface intent for reviewers; to CI/CD pipelines, where agentic workflows~\cite{github2025agent} continuously discover and validate specifications against evolving requirements.
In the age of vibe coding, where humans may never inspect the code, specifications become the \emph{primary interface} between human intent and machine behavior---making this integration not just desirable but essential.

% =============================================================================
\section{Related Work}
\label{sec:related}

Intent formalization draws on and extends several established areas.
\textbf{Specification mining}: classical tools like Daikon~\cite{ernst1999daikon} infer invariants from execution traces but cannot capture user intent beyond observed behavior; LLM-based specification generation transcends this limitation by reasoning about \emph{intended} semantics.
\textbf{LLM-based code generation}: benchmarks like HumanEval~\cite{liu2023evalplus} evaluate code correctness via pass@k but not specification quality; intent formalization shifts the evaluation target from ``does code pass tests?'' to ``do specifications capture intent?''
\textbf{Formal verification}: SMT solvers~\cite{demoura2008z3}, proof assistants, and verification-aware languages (Dafny~\cite{leino2010dafny}, Verus~\cite{lattuada2024verus}) provide mature infrastructure for checking specifications but cannot generate them---intent formalization supplies the missing input.
\textbf{LLMs for verification}: recent work on leveraging LLMs for program verification~\cite{kamath2024fmcad} and Dafny specification synthesis~\cite{misu2024fse} focuses primarily on making verification succeed for given programs rather than capturing user intent for code that does not yet exist.

% =============================================================================
\section{Conclusion}
\label{sec:conclusion}

The age of AI-generated code is here; the age of \emph{reliable} AI-generated code is not.
We have presented early research showing that \textbf{intent formalization}---making user intent explicit, checkable, and enforceable through formal specifications---is a promising direction.
LLMs can generate specifications across the full formalism spectrum.
Automated metrics can evaluate specification quality at or above expert level.
Interactive formalization significantly improves developer correctness on benchmark problems.
End-to-end pipelines produce verified code from informal prose.
And specification quality enables proof automation.

The open challenges---scaling beyond benchmark problems, formalizing change intent, identifying what to clarify cost-effectively, automated metrics for spec validation, handling rich logics, designing human-AI interactions, and achieving compositionality over changes---are each substantial research programs.
We call on the community to treat intent formalization as a \textbf{first-class priority}: with dedicated benchmarks, cross-disciplinary collaboration among AI, PL, formal methods, and HCI researchers, and sustained investment.
The intent gap is the bottleneck; closing it will determine whether AI makes software more reliable or merely more abundant.

\paragraph{Acknowledgments.}
This article reflects contributions from many collaborators, including Nikolaj Bjorner, Sarah Fakhoury, Saikat Chakraborty, Markus Kuppe, Shan Lu, Tahina Ramananandro, Nikhil Swamy, Aaditya Naik, Georgios Sakkas, Madeline Endres, Elizabeth Dinella, Todd Mytkowicz, Madanlal Musuvathi, and many others in the RiSE group at Microsoft Research.
The material presented here draws on work published at FSE 2024~\cite{endres2024fse}, FMCAD 2024~\cite{lahiri2024fmcad}, IEEE TSE 2024~\cite{fakhoury2024tse}, ICSE 2025~\cite{fakhoury2025icse}, ICLR 2025~\cite{chen2025iclr}, SAIV 2025~\cite{sun2025classinvgen}, and TACAS 2026~\cite{sun2026veristruct}.

% --- Bibliography ---
\bibliographystyle{ACM-Reference-Format}
\bibliography{references}

\end{document}